\pdfoutput=1

\documentclass[11pt,a4paper]{article}

\usepackage[T1]{fontenc}
\usepackage[utf8]{inputenc}

\usepackage{mathpazo}

\usepackage[protrusion=true,expansion=true]{microtype}
\usepackage{graphicx}
\usepackage[a4paper,left=1.5in,right=1.5in,top=1.15in,bottom=1.3in]{geometry}
\usepackage{caption}
\usepackage{fancyhdr}
\usepackage{enumitem}
\usepackage[usenames,dvipsnames]{xcolor}
\usepackage[colorlinks=true,allcolors=NavyBlue,breaklinks=true]{hyperref}

\fancypagestyle{plain}{\fancyhf{}%
  \fancyfoot[C]{\footnotesize\thepage}}

\usepackage{titlesec}
\titleformat{\section}{\normalsize\scshape\centering}{}{0pt}{}
\titlespacing*{\section}{0pt}{1.6\baselineskip}{0.5\baselineskip}

\title{Discovery is not the opposite of application}
\author{Hiranya V. Peiris\\
Institute of Astronomy, University of Cambridge, Cambridge, UK\\
Kavli Institute for Cosmology, University of Cambridge, Cambridge, UK}
\date{}

\renewenvironment{abstract}%
  {\begin{center}\begin{minipage}{0.88\textwidth}\large\itshape\noindent}%
  {\end{minipage}\end{center}}

\makeatletter
\renewcommand{\@biblabel}[1]{#1.}
\makeatother

\newcommand{\widefig}[1]{\makebox[\textwidth][c]{\includegraphics[width=1.25\textwidth]{#1}}}

\begin{document}

\thispagestyle{plain}
\begin{center}
{\color{black}\rule{\textwidth}{1.2pt}}\\[0.9em]
{\LARGE Discovery is not the opposite of application\par}
\vspace{0.7em}
{\color{black}\rule{\textwidth}{0.4pt}}\\[1.1em]
{\large\scshape Hiranya V. Peiris\par}
\vspace{0.5em}
{\footnotesize Institute of Astronomy, University of Cambridge, Cambridge, UK\\
Kavli Institute for Cosmology, University of Cambridge, Cambridge, UK\\
\href{mailto:hiranya.peiris@ast.cam.ac.uk}{hiranya.peiris@ast.cam.ac.uk}\par}
\vspace{0.9em}
{\footnotesize\itshape Published as a Comment in Nature Reviews Physics \textbf{8}, 400--401 (2026).\\
Published online 30 June 2026; issue date July 2026.\\
DOI: \href{https://doi.org/10.1038/s42254-026-00964-3}{10.1038/s42254-026-00964-3}\par}
\end{center}

\vspace{1.2em}

\begin{abstract}
Restructuring research funding as a choice between discovery and application gets the economics wrong, the evidence wrong and the people wrong.
\end{abstract}

\vspace{1.4em}

\noindent
A premise is gaining traction in science policy on both sides of the Atlantic: that fundamental and applied research are competing claims on the same budget, and that the balance must shift towards application. In the USA, the current administration has proposed cuts of billions of dollars to the National Institutes of Health and the National Science Foundation. In the UK, the national funding agency UK Research and Innovation (UKRI) is restructuring its portfolio around government growth priorities, with curiosity-driven research nominally protected but the grant lines that actually fund investigators under severe pressure. The language varies --- `doing fewer things better', `aligning research with industrial strategy' --- but the underlying logic is the same. Fundamental science, the argument goes, is a luxury that becomes dispensable when budgets are tight. This premise is wrong, and the consequences of acting on it are already visible.

The economic case for publicly funded fundamental research is well-established. Econometric studies of UK Research Council spending find returns of the order of \href{https://www.universitiesuk.ac.uk/sites/default/files/field/downloads/2024-09/LE-UUK-Impact-of-university-TL-and-RI-Final-Report.pdf}{ten times the investment in GDP} terms~\cite{r1}. A study of the US system, using deliberately conservative assumptions, estimates returns of roughly five dollars per dollar invested~\cite{r2}. These are measurements of past performance, not forecasts. The cost of funding this research is a rounding error on the absolute scale of national expenditure; in the UK, the entire peer-reviewed grant line --- under a billion pounds --- is less than 0.1\% of \href{https://obr.uk/forecasts-in-depth/brief-guides-and-explainers/public-finances/}{total public expenditure}. The cost of losing it is a generation of researchers and their capabilities.

Accepting the premise that fundamental research is dispensable is also a trap. If a country's growth failure is taken as evidence that research investment has not delivered, then no research is safe --- including the applied and mission-driven work that the restructuring claims to prioritize. The UK's \href{https://economy2030.resolutionfoundation.org/}{productivity stagnation} since 2008 has structural causes --- housing, planning, energy costs, chronic underinvestment by business --- that have nothing to do with how grants are allocated between disciplines. Conceding that research must justify itself through near-term contributions to GDP accepts a standard that no scientific enterprise anywhere has ever met. There is also a statistical point: the economic returns from basic research follow a heavy-tailed distribution with divergent variance, dominated by rare transformative discoveries whose timing cannot be predicted~\cite{r3}. Broad cuts across a research base systematically reduce the probability of the rare discoveries that dominate the expected return.

The point is sharpened by the way headline numbers are constructed. UKRI has publicly stated that over half its budget supports curiosity-driven research, but this figure comes from bundling together peer-reviewed grants, block grants to universities, facilities costs, international subscriptions and infrastructure programmes~\cite{r4}. Strip these out, and the share that flows through peer review to investigator-led research is not 50\%. It is 8.8\%. What is counted as curiosity-driven research in these figures bears little resemblance to what working scientists understand by research funding.

A vivid example of why the zero-sum framing of discovery versus application fails is demonstrated by the UK's gravitational wave programme. The UK's Science and Technology Facilities Council (STFC) and its predecessors funded this work for over four decades --- forty years of building precision measurement capability before the first detection of a signal from the merger of two black holes. Forty years of measuring noise better and better --- a `poster-child' of long-horizon science. The Nobel-Prize-winning discovery of gravitational waves in 2015 is an epochal moment in human history --- an entirely new way of sensing the Universe, a capstone in testing general relativity, the very epitome of curiosity-driven research. The University of Glasgow's Institute for Gravitational Research (IGR) made essential contributions towards this discovery in mirror suspension technology, laser stabilization and data analysis.

\begin{figure}[t]
\centering
\widefig{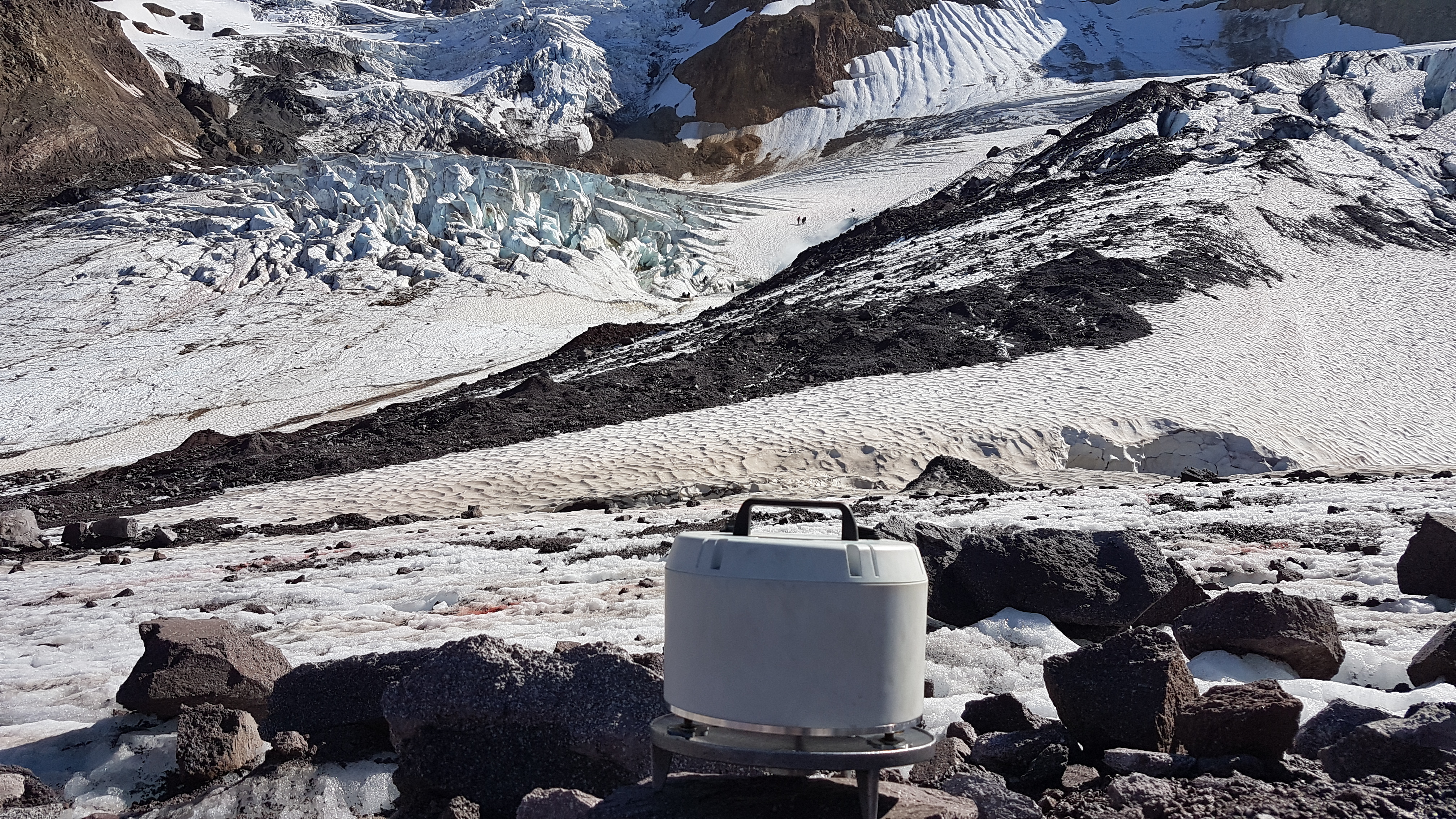}
\caption{\textbf{A Wee-g MEMS gravimeter deployed for field trials on Mount Meager, Canada}. The device --- a direct spin-off from the fused silica suspension technology developed for the LIGO gravitational wave detectors at the University of Glasgow --- measures underground density anomalies with sensitivity a thousand times greater than a mobile phone accelerometer. Applications include defence, GPS-denied navigation, carbon capture site monitoring and geothermal energy exploration. Credit: University of Glasgow.}
\label{fig:weeg}
\end{figure}

The same programme also generated a portfolio of applied innovations --- working technologies, in UK laboratories, produced by the same research groups. A team at Glasgow developed a MEMS gravity sensor --- the Wee-g --- the extraordinary sensitivity of which derives directly from the fused silica suspensions built for the LIGO detectors~\cite{r5}. The device has been deployed on active volcanoes (see Fig.~\ref{fig:weeg}) in multiple countries to monitor sub-surface activity, has attracted funding from defence, energy and security agencies, and was recently incorporated as a spinout company, Quantrologee. Gravitational wave researchers at Strathclyde and Glasgow drew on vibrational measurement expertise from gravitational wave detector work to develop a technique called nanokicking: applying nanoscale vibrations to convert stem cells into bone-building cells without chemicals or scaffolds~\cite{r6}. This work, partly funded by STFC, has gone through NHS clinical investigation for treating disuse osteoporosis following spinal cord injury, and has played a central role in positioning the city of Glasgow as the largest hub for mechanobiology funding in Europe. Other spin-offs from the IGR include Bayesian data analysis techniques now used in retinal imaging for the medical device company Optos, and precision optical coating techniques --- originally developed for gravitational wave detector mirrors --- now in use across quantum technologies, defence and advanced manufacturing. In every case, the discovery and the innovation emerged from the same expertise and the same funding stream. Separate them and you lose both.

Research cultures of this depth take decades to build yet can be dismantled in a single funding cycle. Argentina has experienced repeated cycles of cuts and restoration; each round triggered emigration that was not reversed when funding returned, producing a ratchet of permanent capability loss~\cite{r7}. Portugal's 40\% cut to doctoral fellowships during the eurozone crisis was described by senior scientists as undoing decades of building critical mass~\cite{r8}. The damage was asymmetric in every case: capability destroyed far faster than it could be rebuilt. An open letter signed by over a thousand early-career researchers during the current UK funding crisis framed fundamental science as ``an enabling infrastructure, not a discretionary activity''~\cite{r9}. They are right. The gravitational wave programme took four decades of patient, publicly funded work before it heard its first signal. The expertise that made the discovery possible is the same expertise now entering clinical medicine and volcano monitoring. That is what long-horizon investment in fundamental physics produces, and it cannot be summoned into existence by redirecting funds towards applications that cannot yet be imagined.

The choice being presented to the scientific community --- discovery or application, curiosity or impact --- is a zero-sum illusion. The researchers who pursue fundamental questions are the same ones who end up transforming medicine, defence and industry --- not because they planned to, but because transformative discoveries do not respect funding categories. Fundamental research is the foundation of the building. No one sees it while the skyscraper is standing, but remove it and everything above comes down.

\vspace{0.8em}
\begin{center}{\color{black}\rule{0.25\textwidth}{0.4pt}}\end{center}

\section*{Acknowledgements}
\small\noindent Hiranya Peiris served on STFC Council from 2020 to 2025.

\section*{Competing interests}
\small\noindent The author has received UKRI funding for curiosity-driven research.

\begingroup\small\raggedright

\endgroup

\end{document}